\documentclass[conference]{IEEEtran}
\IEEEoverridecommandlockouts

\usepackage{cite}
\usepackage{amsmath,amssymb,amsfonts}
\usepackage{algorithmic}
\usepackage{graphicx, balance}
\usepackage{textcomp}
\usepackage{xcolor}
\usepackage[font=small,labelfont=bf]{caption}
\usepackage{caption,subcaption}
\usepackage{url}
\def\BibTeX{{\rm B\kern-.05em{\sc i\kern-.025em b}\kern-.08em
    T\kern-.1667em\lower.7ex\hbox{E}\kern-.125emX}}
\begin{document}

\title{ODoQ: Oblivious DNS-over-QUIC\\
}
\author{\IEEEauthorblockN{Aditya Kulkarni\textsuperscript{*}, Tamal Das\textsuperscript{*}, and Vivek Balachandran\textsuperscript{\#}} \\
\IEEEauthorblockA{\textsuperscript{*}\emph{Indian Institute of Technology (IIT) Dharwad, India} \\
\textsuperscript{\#}\emph{Singapore Institute of Technology, Singapore} \\
\textsuperscript{*}\{aditya.kulkarni, tamal\}@iitdh.ac.in, \textsuperscript{\#}vivek.b@singaporetech.edu.sg}}


\maketitle

\begin{abstract}
The Domain Name System (DNS), which converts domain names to their respective IP addresses, has advanced enhancements aimed at safeguarding DNS data and users' identity from attackers. The recent privacy-focused advancements have enabled the IETF to standardize several protocols. Nevertheless, these protocols tend to focus on either strengthening user privacy (like Oblivious DNS and Oblivious DNS-over-HTTPS) or reducing resolution latency (as demonstrated by DNS-over-QUIC). Achieving both within a single protocol remains a key challenge, which we address in this paper. Our proposed protocol -- `Oblivious DNS-over-QUIC' (ODoQ) -- leverages the benefits of the QUIC protocol and incorporates an intermediary proxy server to protect the client's identity from exposure to the recursive resolver.
\end{abstract}


\section{Introduction}
Domain Name System (DNS), often referred to as the ``phone book of the Internet'' is used for translating easy-to-remember domain names (like \texttt{example.com}) into the numerical IP addresses (like \texttt{10.0.2.5}) that network devices use for communication. While humans find it easy to remember names, networked systems rely on IP addresses for connectivity. DNS implements a hierarchical and distributed architecture, which scales efficiently with the growth of the Internet and distributes the load of resolving domain names across globally distributed servers. This design enhances both the reliability and availability of DNS. At the top of this hierarchy are \texttt{root} nameservers, followed by \texttt{Top-Level Domain (TLD)} nameservers, and finally, the \texttt{authoritative} nameservers that provide the final mappings of domain names to IP addresses, ensuring the system’s capacity and resilience in supporting Internet communication.

\begin{figure*}[ht]
    \centering
    \includegraphics[width=1.0\linewidth]{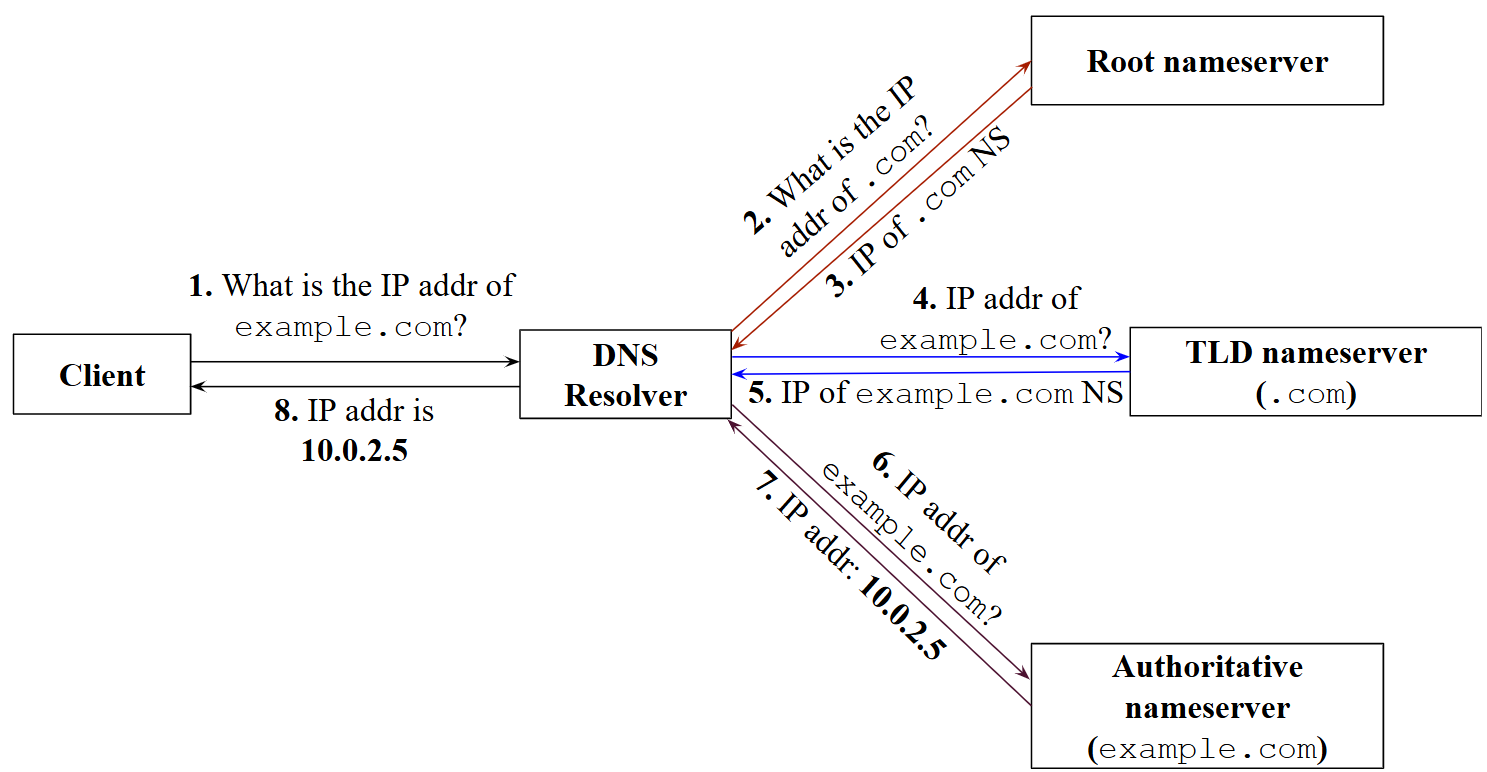}
    \caption{DNS Resolution}
    \label{fig:DNS_Resolution}
\end{figure*}

Figure~\ref{fig:DNS_Resolution} illustrates the DNS resolution for a domain name to retrieve its corresponding IP address. Here a user/client enters a URL (\texttt{example.com}) on a browser's address bar or inputs a command such as \texttt{ping example.com}. The client checks its local cache for the corresponding domain. The client sends a DNS request to a recursive resolver (RR), if the domain name is not found in the local cache. The RR then queries the root servers, the TLD, and authoritative nameservers sequentially to retrieve the IP address associated with the domain name. Upon receiving the DNS response containing the IP address for the requested domain name, the RR caches the information and replies to the client. Finally, the client stores the IP address in its local cache.

Previously, DNS used plain text queries to convert domain names to IP addresses which posed security threats and put users’ privacy at risk due to the lack of encryption on DNS data. The plain text DNS data made it easy for attackers to intercept and track the websites users visit leading to vulnerabilities such as cache poisoning attacks. In these attacks, an attacker injects false records into an RRs cache for a few target domains, thereby redirecting users to phishing websites when queried. Attackers might use the DNS traffic to understand user's behavior by tracking the websites they visit. To overcome these security challenges and safeguard the DNS data from eavesdroppers, several protocols have been proposed and standardized by the Internet Engineering Task Force (IETF)\footnote{\url{https://www.ietf.org/}}. These security measures consist of protocols such, as DNS Security Extension (DNSSEC)~\cite{DNSSEC_RFC}, DNS-over-TLS (DoT)~\cite{DoT_RFC}, DNS-over-DTLS (DoDTLS)~\cite{DoDTLS_RFC}, DNS-over-HTTPS (DoH)~\cite{DoH_RFC}, Oblivious DNS (ODNS)~\cite{schmitt2018oblivious}, Oblivious DoH (ODoH)~\cite{ODoH_RFC}, and DNS-over-QUIC (DoQ)~\cite{DoQ_RFC}. To enhance privacy, RRs use \texttt{QNAME} minimization~\cite{QNAME_RFC} to only disclose parts of domain names during resolution to nameservers, as illustrated in Figure~\ref{fig:DNS_Resolution}.

The IETF has standardized DoT, DoH, and DoQ protocols as secure alternatives for plain text DNS. These protocols secure communication between the client and the RR. However, the RR continues to possess information about the client’s identity, which creates possible privacy concerns. If the RR gets compromised, it might utilize the client's identity to monitor browsing behavior or even misuse this data for target advertisements. To enhance user privacy, Oblivious DNS (ODNS)\cite{schmitt2018oblivious} and Oblivious DoH (ODoH)\cite{ODoH_RFC} have been proposed, which use an intermediary proxy server to hide the client's identity from getting exposed to the RR. The client uses the RR's public key to encrypt the DNS query ensuring that it gets decrypted only by the RR. The proxy knows only the IP address of the client and does not know the domain it is querying, as the proxy cannot decrypt it. Whereas, the RR knows only the domain name, but not the IP address of the client, as the proxy removes the client's IP address and transmits the packet containing the encrypted DNS request to the RR for resolution. The proxy and RR must be managed by different organizations to avoid the correlation of DNS messages, which could lead to the identification of the client.

A recent study~\cite{kosek2022dns} found that DoQ is roughly $10\%$ faster than DoH, and despite adding some encryption overhead, DoQ is merely $2\%$ slower compared to plain text DNS over UDP. This renders DoQ a very effective and trustworthy protocol for sending DNS messages between the client and the RR. Nonetheless, a significant drawback of DoQ is that it fails to preserve the client's identity from the RR. This privacy concern drives us to suggest an enhancement to the DoQ protocol that includes a proxy server as an intermediary, which would preserve the client's anonymity by concealing the client's identity from the RR.

\hfill

\noindent \textit{Our Contribution}: In this paper, we propose a more secure protocol, Oblivious DNS-over-QUIC (ODoQ), aimed to preserve the client's privacy from the RR. ODoQ utilizes the benefits of the QUIC protocol to establish a secure connection between the client and the proxy, as well as between the proxy and the RR. We provide a comprehensive overview of the DNS resolution process using ODoQ and also discuss the steps taken if there is a \textit{decryption failure} at the RR.

\hfill

\noindent \textit{Organization of the paper}: The remainder of this paper is as follows. Section~\ref{sec:Related_Work} describes various protocols that add security and privacy implications to the plain text DNS. In Section~\ref{sec:ODoQ_Protocol}, we describe our proposed protocol ODoQ to enhance the privacy of the client's identity. Finally in Section~\ref{sec:Conclusion_and_Future_Directions}, we conclude with future directions on our proposed ODoQ protocol.

\section{Related Work}
\label{sec:Related_Work}
In the early days, DNS operated using plain text queries to translate domain names to IP addresses. The lack of privacy in this process exposed users' browsing patterns, including the websites they visited, which could allow attackers to monitor their behavior. This vulnerability could also lead to attacks such as redirecting users to phishing websites designed to steal users' sensitive information. As illustrated in Figure~\ref{fig:DNS_Resolution}, an attacker could exploit DNS traffic between the client and RR, or between RR and the nameservers. DNS communication between the client and the RR can reveal the client's IP address and the queried domain name, providing enough information for an attacker to profile the client's online activity. However, in the DNS resolution process between the RR and the nameservers, only the domain name is exposed.

\begin{figure*}
    \centering
    \includegraphics[width=1.0\linewidth]{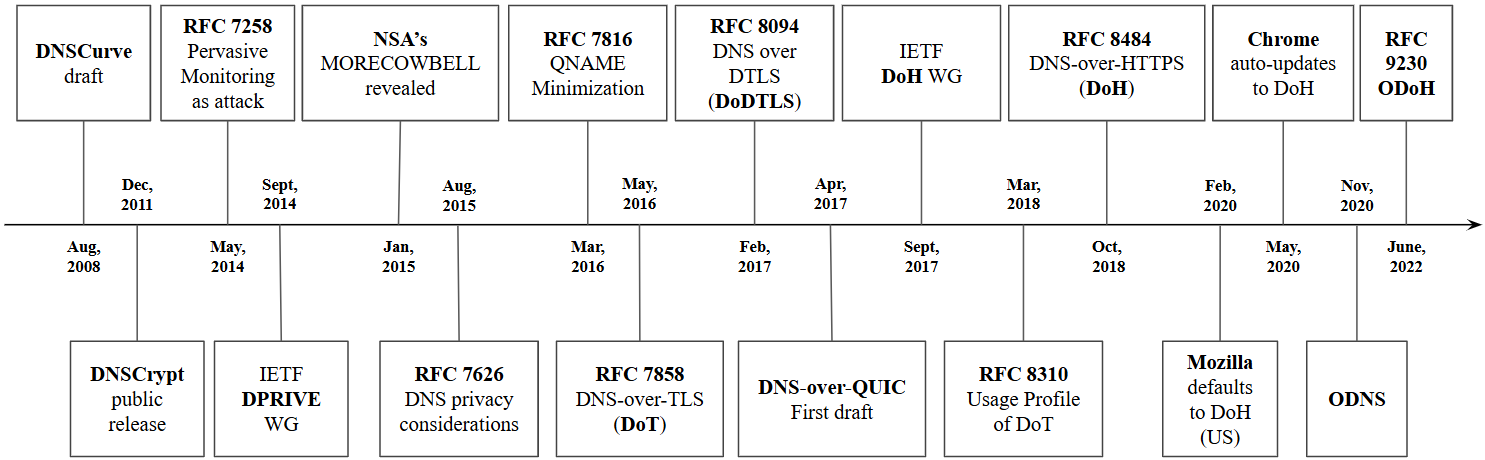}
    \caption{DNS Evolution}
    \label{fig:DNS_Evolution}
\end{figure*}

Figure~\ref{fig:DNS_Evolution} charts the DNS evolution. Several protocols have been proposed over the past decade to secure the DNS data, and to preserve the client's information (such as the IP address and the domain name) during the client-RR communication. These protocols that secure the DNS data between the client and the RR include DoT~\cite{DoT_RFC}, DoH~\cite{DoH_RFC}, DoDTLS~\cite{DoDTLS_RFC} (an experimental protocol that uses UDP for transport, unlike DoT which uses TCP) and DoQ. On the other hand, ODNS~\cite{schmitt2018oblivious} and ODoH~\cite{ODoH_RFC} protocols focus on preserving the client's privacy from the RR by introducing an intermediary proxy server.

DNS-over-TLS (DoT)~\cite{DoT_RFC} was standardized by the IETF in $2015$ under RFC $7858$. In this protocol, the client establishes a TCP three-way handshake with the DNS RR, followed by a TLS handshake. DoT messages are typically transmitted over port $853$, although either party may agree to use a different port if preferred. The chosen port must vary from port $53$ to avoid conflicts, minimizing DNS interference and lowering the risk of downgrade attacks. Given that DoT employs a comparable message format to standard plain text DNS over UDP, it is quite easy to implement as a secure alternative.

DNS-over-HTTPS (DoH)~\cite{DoH_RFC} was standardized by the IETF in $2017$ under RFC $8484$. In this protocol, the client first establishes a three-way TCP handshake with the DNS RR, and then encapsulates the DNS query within an HTTP packet, which is transmitted to the RR. DoH offers a degree of DNS query encryption comparable to the DoT, but utilizes port $443$ for its communication. The benefit of utilizing port $443$ is that an eavesdropper cannot differentiate between a standard HTTPS packet and a DoH packet.

Both DoT and DoH protect privacy by employing the proven encryption methods of TLS, although this results in extra delays when resolving DNS queries. The implementation of TCP contributes to latency (due of its three-way handshake) and head-of-line blocking (since it requires packet ordering). Furthermore, TLS adds its own processing and setup time for DNS resolution. The DoQ~\cite{DoQ_RFC} protocol tackles these overheads by utilizing QUIC (Quick UDP Internet Connections), a transport layer protocol based on UDP created by Google. Originally, gQUIC was created to enhance video streaming performance and was subsequently accepted and formalized by the IETF as RFC 9000~\cite{gQUIC_RFC}. Afterwards, the QUIC protocol was modified to accommodate various protocols, such as DNS, and was formalized by the IETF as RFC 9001~\cite{QUIC_RFC}, using TLS v1.3. Additionally, QUIC enables the multiplexing of packets over various streams within a single connection, thereby eliminating the head-of-line blocking problem found in TCP. DoQ~\cite{DoQ_RFC} leverages the QUIC protocol to enhance the security and privacy of DNS packets by placing them inside QUIC packets.

As the use of encrypted DNS and public DNS resolvers to improve privacy increases, a significant issue emerges: although clients' DNS information is secured from unauthorized access, the RR still retains access to the clients' identities. This prompts one to consider whether it is wiser to rely on an external provider—typically situated in another country with a distinct legal framework—than on the local Internet Service Provider (ISP) of the client. To prevent the RR from acquiring full knowledge of the client's IP address and domain name, protocols such as ODNS~\cite{schmitt2018oblivious} and ODoH~\cite{ODoH_RFC} utilize a proxy server that serves as an intermediary between the client and the RR to preserve client's identity from getting revealed to the RR.

The primary objective of the proxy in the ODNS and ODoH protocols is to stop the client from directly connecting with the RR and revealing its identity. Instead, the proxy server serves as an intermediary, hiding the client's identity from the RR. Initially, the client creates a DNS query and encrypts it using the RR's public key, ensuring that only the RR can decrypt it with its private key. The client establishes a secure connection with the proxy, encapsulates the encrypted DNS request in an HTTP packet, and forwards it to the proxy. The proxy is aware of the client's IP address but is unable to decrypt or examine the requested domain. Upon obtaining the packet, the proxy extracts the RR's address from the metadata within the received packet and initiates a secure connection with the RR. It subsequently encapsulates the encrypted DNS request within another HTTP packet and sends it to the RR. The RR uses its associated private key to decrypt the request and processes it to obtain the associated IP address. The RR is unaware of the client's IP address. The RR then encrypts the DNS response with the client's encryption key, encapsulates it in an HTTP packet, and returns it to the proxy. Ultimately, the proxy sends the encrypted DNS response to the client, who can decrypt it with its key.

\section{Proposed Approach: \\ ODoQ -- Oblivious DNS-over-QUIC}
\label{sec:ODoQ_Protocol}
In this section, we describe our proposal, ``Oblivious DNS-over-QUIC (ODoQ)'', which leverages the benefits of the QUIC protocol, similar to DoQ, and introduces an intermediary proxy server to ensure that the client's privacy is preserved from the RR. ODoQ enables DNS messages to be encapsulated within HTTP packets transmitted over a secure QUIC connection between the client and the proxy, as well as between the proxy and the RR.

\begin{figure*}[ht]
    \centering
    \includegraphics[width=1.0\linewidth]{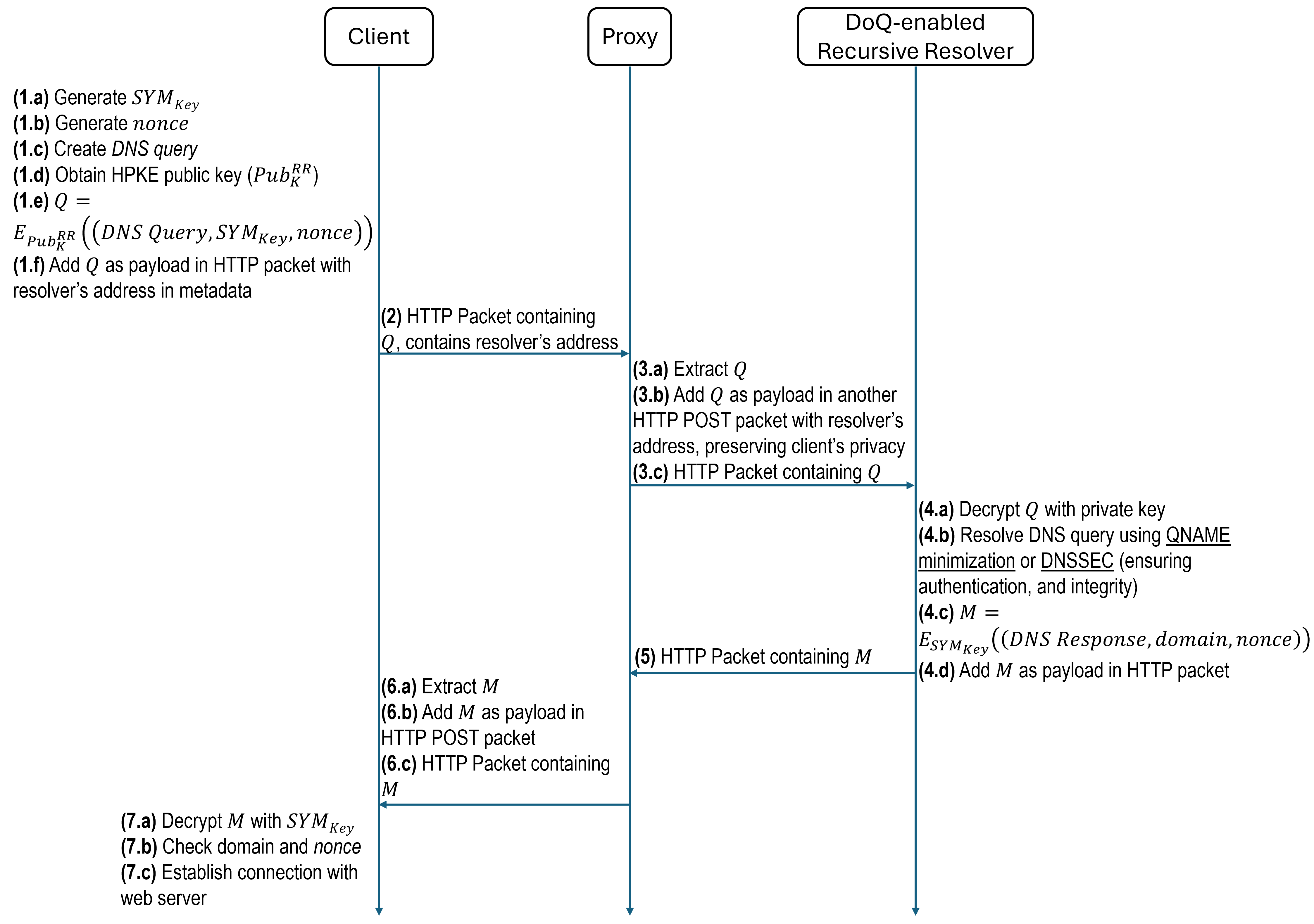}
    \caption{ODoQ protocol}
    \label{fig:ODoQ_Protocol}
\end{figure*}

The three entities used in the ODoQ protocol are as follows:
\begin{enumerate}
    \item \textit{Client}: the end system that initiates the DNS query and upon receiving the associated IP address for the domain will establish a secure connection with the associated web server.
    \item \textit{Proxy Server}: an intermediary that forwards the DNS queries from the client to the DoQ-enabled RR, ensuring that the client's identity remains hidden from the RR. Also, it forwards the response from the DoQ-enabled RR back to the client, which initiated the DNS request.
    \item \textit{DoQ-enabled RR}: A RR that receives the DNS query (requested by the client) from the proxy server, decrypts, resolves and sends the response back to the proxy server.
\end{enumerate}

Unlike direct resolution in DoT, DoH, and DoQ, in our ODoQ protocol, the HTTP exchange happens between the three entities which maintains the privacy of the DNS data and also preserves the client's privacy. Figure~\ref{fig:ODoQ_Protocol} shows the DNS query resolution in ODoQ protocol that uses the proxy to preserve the client's privacy from the RR.
\begin{enumerate}
    \item \textit{Client initiates DNS query using Hybrid Public Key Encryption (HPKE~\cite{HPKE_RFC})}
    \begin{itemize}
        \item The client creates a DNS query to know the corresponding IP address for connecting with the intended web server.
        \item It generates its symmetric key $SYM_{Key}$, and a $nonce$. the RR will later use the $SYM_{Key}$ to encrypt the DNS response and transmit it to the client through the proxy server, ensuring that only the client can decrypt the DNS response. The $nonce$ is a unique number to avoid replay attacks.
        \item The client obtains the HPKE public key $Pub_K^{RR}$ of the RR using some pre-shared mechanism and encrypts the DNS query, $SYM_{Key}$, and the $nonce$, thus generating an \textit{encrypted DNS request}:  \\ 
        $ Q = E_{Pub_K^{RR}}((DNS \text{ } Query, \text{ } SYM_{Key}, \text{ } nonce))$
    \end{itemize}
    \item \textit{Client establishes QUIC connection with the proxy}
    \begin{itemize}
        \item The client establishes a secure QUIC connection with the proxy.
        \item It adds the encrypted DNS request ($Q$) as a payload into an HTTP POST packet. It provides the address (a destination URI) of the RR as metadata in the packet so that the proxy can establish a secure connection. Since the client has used the resolver's HPKE public key to encrypt, it can only be decrypted with the associated private key.
    \end{itemize}
    \item \textit{Proxy receives and forwards $Q$ to the RR}
    \begin{itemize}
        \item Upon receiving the packet from the client, the proxy extracts the $Q$, the resolver's address present in the metadata and establishes a secure connection with the RR.
        \item The proxy constructs another HTTP packet containing $Q$ as a payload and forwards it to the resolver, thereby preserving the client's privacy.
    \end{itemize}
    \item \textit{Resolver receives $Q$, decrypts it, and processes the DNS query}
    \begin{itemize}
        \item The RR receives the packet from the proxy, extracts $Q$, and decrypts it using its associated private key to obtain the DNS query, the client's symmetric key $SYM_{Key}$, and the client's $nonce$.
        \item To enhance privacy, the resolver can use \texttt{QNAME minimization} or \texttt{DNSSEC} to resolve the DNS query and get the associated IP address from an authoritative nameserver. Also, this will not reveal the full query to the hierarchy nameservers (root, and TLD).
    \end{itemize}
    \item \textit{Recursive resolver resolves DNS query and encrypts the response}
    \begin{itemize}
        \item The resolver resolves the DNS query to obtain the associated IP address.
        \item It encrypts the DNS response, the domain name and the client's $nonce$ using the client's symmetric key $SYM_{Key}$, ensuring that only the client can decrypt it, i.e., \\ 
        $M = E_{SYM_{Key}}((DNS \text{ } response, domain, nonce))$.
        \item The resolver adds $M$ as a payload inside an HTTP packet and transmits it to the proxy.
    \end{itemize}
    \item \textit{Proxy receives $M$ and forwards it to client}
    \begin{itemize}
        \item The proxy receives the HTTP packet from the resolver, and extracts $M$ (encrypted DNS response).
        \item It creates another HTTP packet, adds $M$ as payload in it and transmits it to the client.
    \end{itemize}
    \item \textit{Client receives $M$, decrypts it, and verifies the response}
    \begin{itemize}
        \item The client on receiving the packet from the proxy, extracts the $M$ (encrypted DNS response) and decrypts it using $SYM_{Key}$.
        \item It verifies the response integrity by checking if the domain name and the $nonce$ in the response match the one originally requested.
        \item If the domain name and the $nonce$ match, the client ensures that the DNS resolution is correct and the packet has not been tampered with during the resolution.
    \end{itemize}
    \item \textit{Client establishes a secure connection with web server}
    \begin{itemize}
        \item The client establishes a secure connection with the web server using the DNS response IP address.
        \item The connection can be over a TLS 1.3 handshake using encrypted Server Name Indication (ESNI)~\cite{ESNI_RFC} or Encrypted Client Hello (ECH)~\cite{ECH_RFC}.
    \end{itemize}
\end{enumerate}

A recent study~\cite{kosek2022one} performed a comprehensive study to assess the performance, robustness and deployment of the DoQ protocol in the wild. The study findings reveal that the DoQ adoption in the RRs is still restricted and only AdGuard\footnote{\url{https://adguard-dns.io/kb/general/dns-providers/}} is one of the first public providers to adopt and provide DoQ functionality. These findings help implement our ODoQ protocol and observe its performance with the existing DoT, DoH, DoQ, ODNS and ODoH protocols. We believe that the two-fold strategy of our ODoQ protocol -- preserving client's privacy and encrypted communication over QUIC -- increases the privacy and security of DNS data.

ODoQ seeks to maintain the client's privacy by utilizing a proxy that is prevented from inspecting the domain name the client wishes to connect to, as the DNS request is secured using the RR's public key. This raises a question of \textit{decryption failure}, where the RR is unable to decrypt the encrypted DNS request due to a possible update of its public-private key pair. One way to address this issue is for the RR to transmit the new public key along with an error message -- indicating the failure to decrypt the prior DNS request -- to the client through the proxy on the existing connection, preventing the establishment of a new connection and raising the overheads. After the client receives the updated key and realizes that its DNS request was not fulfilled because of decryption issues. The client refreshes itself with the updated public key of the RR and utilizes the same QUIC connection to continue with the identical message transfer for DNS query resolution.

\begin{figure}[htpb]
    \centering
    \begin{subfigure}{0.45\textwidth}
        \centering
        \includegraphics[width=1.0\textwidth]{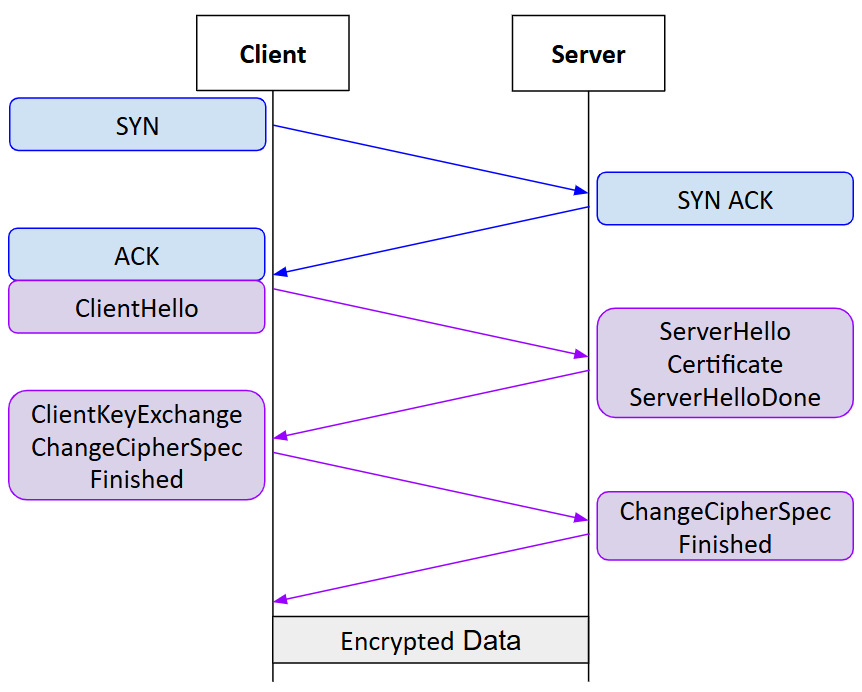}
        \caption{TCP + TLS 1.2}
        \label{fig:TCP_TLS_1_2}
    \end{subfigure}
    \vspace{0.2em}
    \begin{subfigure}{0.45\textwidth}
        \centering
        \includegraphics[width=1.0\textwidth]{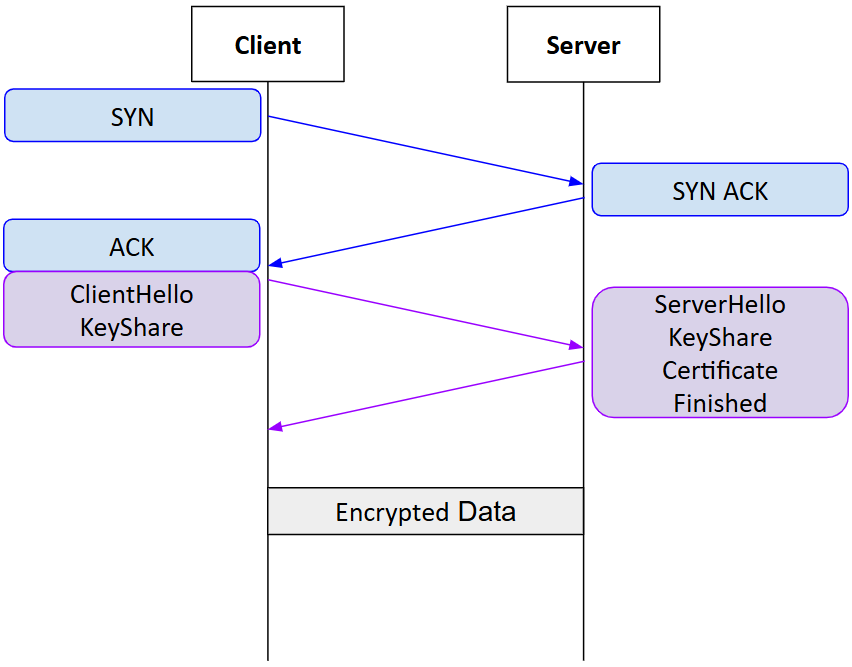}
        \caption{TCP + TLS 1.3}
        \label{fig:TCP_TLS_1_3}
    \end{subfigure}
    \vspace{0.2em}
    \begin{subfigure}{0.45\textwidth}
        \centering
        \includegraphics[width=1.0\textwidth]{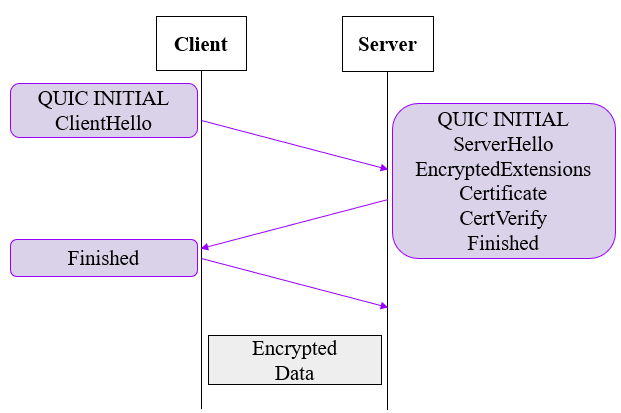}
        \caption{QUIC}
        \label{fig:QUIC}
    \end{subfigure}
    \caption{Handshake process of (\ref{fig:TCP_TLS_1_2}) TCP + TLS 1.2, (\ref{fig:TCP_TLS_1_3}) TCP + TLS 1.3, and (\ref{fig:QUIC}) QUIC connection.}
    \label{fig:Handshakes}
\end{figure}

Figure~\ref{fig:Handshakes} illustrates the round-trip times (RTTs) required to establish a connection between the client and the server for various protocols. For TCP + TLS 1.2 (Figure~\ref{fig:TCP_TLS_1_2}), the process requires 1 RTT for TCP and 2 RTTs for TLS 1.2. TCP + TLS 1.3 (Figure~\ref{fig:TCP_TLS_1_3}) reduces this to 1 RTT for TCP and 1 RTT for TLS 1.3. Finally, QUIC (Figure~\ref{fig:QUIC}) requires 1 RTT for QUIC protocol. DoT and DoH require 3 RTTs using TLS 1.2, involving one RTT for the TCP handshake and two for the TLS handshake. TLS 1.3 cuts this down to 2 RTTs, one for TCP and one for TLS. DoQ has a quicker process, requiring just 1 RTT for the handshake because of QUIC's combined transport and encryption. ODoH introduces an intermediary proxy server between the client and the resolver, causing communication between client-proxy and proxy-resolver to require 2 RTTs along with an extra overhead from HTTP/2. ODoQ utilizes a proxy alongside QUIC to decrease latency to only 2 RTTs required for handshakes among the client, proxy, and resolver.

\section{Conclusion and Future Directions}
\label{sec:Conclusion_and_Future_Directions}
In this paper, we proposed the Oblivious DNS-over-QUIC (ODoQ) protocol, utilizing the benefits of the QUIC protocol in conjunction with an intermediary proxy server to safeguard client privacy from the recursive resolver (RR). The paper describes the secure message exchange among the client, proxy, and RR, in which DNS data is encrypted both ways -- client to RR and from RR to client -- utilizing the RR’s public key and the client’s symmetric key correspondingly. We also address measures to manage decryption failure at the RR in situations where key pairs have been updated. As a future direction, we intend to create a prototype of the ODoQ protocol utilizing existing tools and public resolvers that facilitate DoQ. In addition, we aim to study and compare the overhead and performance of ODoQ with the existing DNS over encryption protocols.

\balance
\bibliographystyle{IEEEtran}
\bibliography{references}

@misc{DoT_RFC,
    title = {{Specification for DNS over Transport Layer Security (TLS)}},	
    author = {Zi, Hu and Liang, Zhu and John, Heidemann and Allison, Mankin and Duane, Wessels and Paul, E. Hoffman},
    year = {2016},   
    howpublished = {RFC 7858, \url{https://datatracker.ietf.org/doc/html/rfc7858} [Accessed: November 8th, 2024]},
}

@misc{DoDTLS_RFC,
    title = {{DNS over Datagram Transport Layer Security (DTLS)}},	
    author = {Tirumaleswar, Reddy.K and Dan, Wing and Prashanth, Patil},
    year = {2017},
    howpublished = {RFC 8094, \url{https://datatracker.ietf.org/doc/html/rfc8094} [Accessed: November 8th, 2024]},
}

@misc{DoH_RFC,
    title = {{DNS Queries over HTTPS (DoH)}},	
    author = {Paul, E. Hoffman and Patrick, McManus},
    year = {2018},
    howpublished = {RFC 8484, \url{https://datatracker.ietf.org/doc/html/rfc8484} [Accessed: November 8th, 2024]},
}

@misc{DNSSEC_RFC,
    title = {{DNS Security Extensions (DNSSEC)}},	
    author = {Paul, E. Hoffman},
    year = {2023},
    howpublished = {RFC 9364, \url{https://datatracker.ietf.org/doc/html/rfc9364} [Accessed: November 8th, 2024]},
}

@article{schmitt2018oblivious,
  title={{Oblivious DNS: Practical Privacy for DNS Queries}},
  author={Schmitt, Paul and Edmundson, Anne and Feamster, Nick},
  journal={arXiv preprint arXiv:1806.00276},
  year={2018}
}

@misc{ODoH_RFC,
    title = {{Oblivious DNS over HTTPS}},	
    author = {Eric, Kinnear and Patrick, McManus and Tommy, Pauly and Tanya, Verma and Christopher, A. Wood },
    year = {2022},
    howpublished = {RFC 9230, \url{https://datatracker.ietf.org/doc/html/rfc9230} [Accessed: November 8th, 2024]},
}

@misc{DoQ_RFC,
    title = {{DNS over Dedicated QUIC Connections}},	
    author = {Christian, Huitema and Sara, Dickinson and Allison, Mankin},
    year = {2022},
    howpublished = {RFC 9250, \url{https://datatracker.ietf.org/doc/rfc9250/} [Accessed: November 8th, 2024]},
}

@misc{QNAME_RFC,
    title = {{DNS Query Name Minimisation to Improve Privacy}},	
    author = {Stéphane, Bortzmeyer and Ralph, Dolmans and Paul, E. Hoffman},
    year = {2021},
    howpublished = {RFC 9156, \url{https://datatracker.ietf.org/doc/rfc9156/} [Accessed: November 8th, 2024]},
}

@inproceedings{kosek2022dns,
  title={{DNS Privacy with Speed? Evaluating DNS over QUIC and its Impact on Web Performance}},
  author={Kosek, Mike and Schumann, Luca and Marx, Robin and Doan, Trinh Viet and Bajpai, Vaibhav},
  booktitle={Proceedings of the 22nd ACM Internet Measurement Conference},
  pages={44--50},
  year={2022}
}

@misc{gQUIC_RFC,
    title = {{QUIC: A UDP-Based Multiplexed and Secure Transport}},	
    author = {Jana, Iyengar and Martin, Thomson},
    year = {2021},
    howpublished = {RFC 9000, \url{https://datatracker.ietf.org/doc/html/rfc9000} [Accessed: November 8th, 2024]},
}

@misc{QUIC_RFC,
    title = {{Using TLS to Secure QUIC}},	
    author = {Martin, Thomson and Sean, Turner},
    year = {2021},
    howpublished = {RFC 9001, \url{https://datatracker.ietf.org/doc/html/rfc9001} [Accessed: November 8th, 2024]},
}

@inproceedings{kosek2022one,
  title={{One to Rule them All? A First Look at DNS over QUIC}},
  author={Kosek, Mike and Doan, Trinh Viet and Granderath, Malte and Bajpai, Vaibhav},
  booktitle={International Conference on Passive and Active Network Measurement},
  pages={537--551},
  year={2022},
  organization={Springer}
}

@misc{ESNI_RFC,
    title = {{Encrypted Server Name Indication for TLS 1.3}},	
    author = {Eric, Rescorla and Kazuho, Oku and Nick, Sullivan and Christopher A., Wood},
    year = {2019},
    howpublished = {IETF Draft, \url{https://datatracker.ietf.org/doc/html/draft-ietf-tls-esni-02} [Accessed: November 11th, 2024]},
}

@misc{ECH_RFC,
    title = {{TLS Encrypted Client Hello}},	
    author = {Eric, Rescorla and Kazuho, Oku and Nick, Sullivan and Christopher A. Wood},
    year = {2024},
    howpublished = {IETF Draft, \url{https://datatracker.ietf.org/doc/draft-ietf-tls-esni/} [Accessed: November 11th, 2024]},
}

@misc{HPKE_RFC,
    title = {{Hybrid Public Key Encryption}},	
    author = {Richard, Barnes and Karthikeyan, Bhargavan and Benjamin, Lipp and Christopher A., Wood},
    year = {2022},
    howpublished = {RFC 9180, \url{https://datatracker.ietf.org/doc/rfc9180/} [Accessed: November 11th, 2024]},
}

\end{document}